\documentclass[pdflatex,sn-basic]{sn-jnl}

\usepackage{graphicx}%
\usepackage{multirow}%
\usepackage{amsmath,amssymb,amsfonts}%
\usepackage{amsthm}%
\usepackage{mathrsfs}%
\usepackage[title]{appendix}%
\usepackage{xcolor}%
\usepackage{textcomp}%
\usepackage{manyfoot}%
\usepackage{booktabs}%
\usepackage{algorithm}%
\usepackage{algorithmicx}%
\usepackage{algpseudocode}%
\usepackage{listings}%

\usepackage{wasysym}

\theoremstyle{thmstyleone}%
\theoremstyle{thmstyletwo}%

\theoremstyle{thmstylethree}%

\begin{document}

\title[Formation of super-Jupiters]{On the formation of super-Jupiters: Core Accretion or Gravitational Instability?}


\author*[1]{\fnm{Max} \sur{Nguyen}}\email{MaxHungNguyen@gmail.com}

\author[2]{\fnm{Vardan} \sur{Adibekyan}}\email{vadibekyan@astro.up.pt}

\affil*[1]{Leland High School, 6677 Camden Avenue, San Jose, CA 95120, USA}

\affil[2]{ Instituto de Astrof\'isica e Ci\^encias do Espa\c{c}o, Universidade do Porto, CAUP, Rua das Estrelas, 4150-762 Porto, Portugal}


\abstract{The Core Accretion model is widely accepted as the primary mechanism for forming planets up to a few Jupiter masses. However, the formation of super-massive planets remains a subject of debate, as their formation via the Core Accretion model requires super-solar metallicities. Assuming stellar atmospheric abundances reflect the composition of protoplanetary disks, and that disk mass scales linearly with stellar mass, we calculated the total amount of metals in planet-building materials that could contribute to the formation of massive planets. In this work, we studied a sample of 172 Jupiter-mass planets and 93 planets with masses exceeding 4 $M_{\jupiter}$.Our results consistently demonstrate that planets with masses above 4 $M_{\jupiter}$ form in disks with at least as much metal content as those hosting planets with masses between 1 and 4 $M_{\jupiter}$, often with slightly higher metallicity, typically exceeding that of the proto-solar disk. We interpret this as strong evidence that the formation of very massive Jupiters is feasible through Core Accretion and encourage planet formation modelers to test our observational conclusions.}


\maketitle

\section{Introduction} \label{sec:intro}

There are two widely discussed models for the formation of massive planets: Core Accretion (CA) \citep{Pollack-96} and Gravitational Instability (GI) \citep{Boss-97}. The well-established correlation between giant planets and host star metallicity \citep{Santos-01, Fischer-05} supports CA as the primary formation mechanism for these planets. However, this correlation appears to hold primarily for planets with masses up to approximately 4 $M_{\jupiter}$. \citet{Adibekyan-13} examined the metallicity distributions of stars hosting planets with masses between 1 and 4 $M_{\jupiter}$ (hereafter referred to as Jupiters) and those hosting more massive planets between 4 and about 13 $M_{\jupiter}$ (hereafter referred to as super-Jupiters). Although their analysis showed that the hosts of super-Jupiters tend to be slightly less metallic, the difference was not statistically significant.

\citet{Santos-17a} found that stars hosting Jupiters generally have higher metallicities than field stars (stars without planets) of similar mass. In contrast, the hosts of super-Jupiters share a similar metallicity distribution with field stars of comparable mass. The authors suggested that Jupiters likely form via CA, whereas super-Jupiters may form through GI. In a subsequent study, \citet{Adibekyan-19} found that giant planets both above and below 4 $M_{\jupiter}$ orbiting solar-type dwarfs are preferentially metal-rich. However, \citet{Adibekyan-19} observed differences in the metallicity distribution among host stars with masses greater than 1.5 $M_{\odot}$, leading to a proposal that the planet formation mechanism is influenced more by environmental factors, such as metallicity and disk mass, rather than planet mass alone.

Interestingly, based on the assumption that sub-stellar objects can be distinguished by the metallicity of their host stars, \citet{Schlaufman-18} proposed that the boundary between objects formed through CA and GI lies between 4 and 10 $M_{\jupiter}$.

The aforementioned results are somewhat contradictory, highlighting the complexity of understanding how super-Jupiters form \citep[see also][]{Narang-18, Maldonado-19}. One key limitation of these studies is the use of iron abundance ([Fe/H]) as a proxy for overall metallicity, which may not fully capture the true metal content of the protoplanetary disks. This simplification could obscure the underlying mechanisms of planet formation. As a result, there is a clear need to address this open question and further investigate how super-massive planets form, taking into account a more comprehensive estimate of the total metal content in protoplanetary disks. In this manuscript, we aim to contribute to this effort by considering the total metal content in the disks.

\section{Sample} \label{sec:sample}

We began by selecting planets with masses between 1 and 13 Jupiter masses from the NASA Exoplanet Archive (NEA, \href{https://catcopy.ipac.caltech.edu/dois/doi.php?id=10.26133/NEA12}{DOI:10.26133/NEA12}). The upper limit of 13 M$_{jup}$ was chosen as it approximately corresponds to the lower mass boundary for brown dwarfs \citep{Spiegel-11}. Out of a total of 5747 confirmed planets in the NEA, we selected 483 planets with Radial Velocity measurements and with relative uncertainties on mass of less than 50\%\footnote{We considered the 'pl\_bmassj' which is the best planet mass estimate available at NEA, in order of preference: Mass,$M \sin(i)/\sin(i)$, or$M \sin(i)$, depending on availability.}. 

To obtain stellar parameters and host star masses, we cross-matched the list of host stars with the SWEET-Cat database \citep{Santos-13, Sousa-24}\footnote{http://sweetcat.iastro.pt/}. This resulted in a sample of 428 massive planets orbiting 396 stars.

To obtain the stellar abundances of the host stars, we cross-matched this list with the Hypatia Catalog \citep{Hinkel-14}, a compilation of stellar abundances for nearby stars. We focused on host stars with available abundances of C, O, Mg, Si, and Fe, as these elements are crucial for determining the metal content of planet-building materials. As evident from the APOGEE data, the Galaxy has not yet enriched in metals beyond [X/H] $\sim$ 0.6 dex, where X represents one of the elements we considered \citep[e.g.][]{Ratcliffe-23}. Therefore, we selected only stars with [X/H] $<$ 0.6, which allowed us to exclude four targets with potentially inaccurate abundances, such as HIP\,88414 with [C/H] = 1.41 dex.

Our final sample consists of 172 Jupiter-mass planets orbiting 152 stars, along with 93 super-Jupiters orbiting 88 host stars. Their distribution on the  $M_{\mathrm{pl}}$ - $M_{\mathrm{star}}$ and $M_{\mathrm{pl}}$ - [Fe/H] is displayed on Figure 1. The hosts cover a wide range of metallicity, stellar masses, and evolutionary stages (1.76 $<$ $\log g$ $<$ 4.8 dex).

\begin{figure}[]
\centering
\includegraphics[width = 1\textwidth]{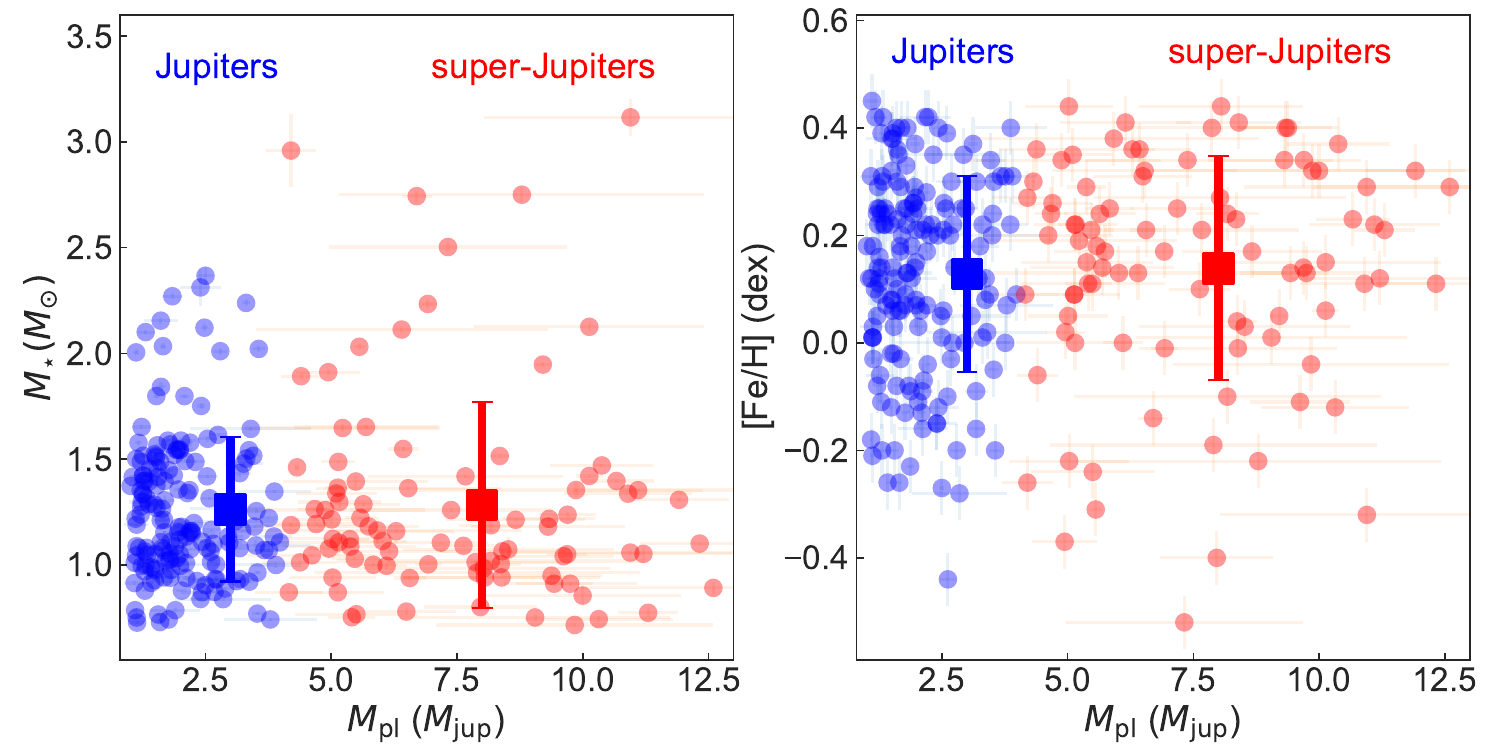}
\caption{Planetary masses as a function of host star mass (left panel) and host star metallicity (right panel). The blue and red small circles represent Jupiters and super-Jupiters, respectively. The large blue and red squares indicate the mean and standard deviation for Jupiters and super-Jupiters, respectively.}\label{fig:1}
\end{figure}

The abundances in the Hypatia Catalog are compiled from various sources and are not derived homogeneously. We compared the iron abundances from the Hypatia Catalog with those determined homogenously in the SWEET-Cat database. We selected only the abundances with homogeneous determinations by setting the SWFlag to 1 in the SWEET-Cat database. The mean difference between the two databases is -0.03 $\pm$ 0.05 dex, which is comparable to the mean uncertainty of [Fe/H] in the SWEET-Cat database (0.03 dex).

\section{Metals in the protoplanetary disk} \label{sec:metals}

We used the solar reference values from \citet{Asplund-21} to convert the relative abundances (C, O, Mg, Si, and Fe) of the host stars into absolute abundances. We then used these abundances to estimate the total mass fraction of heavy elements ($Z$) of the planet forming blocks, following the stoichiometric model presented in \citet{Santos-17b}. Here we assume that the stellar atmospheric composition reflects the composition of the protoplanetary disk.  Based on the solar abundances from \citet{Asplund-21}, the model predicts a $Z_{\odot}$ of 1.25\% for the planet-building blocks in the Solar System. Since the Hypatia Catalog does not provide uncertainties for the abundances, we adopted the median uncertainties from a sample of $\sim$1100 stars observed with high-resolution HARPS spectra: $\sigma_{C} = 0.09$ dex, $\sigma_{O} = 0.11$ dex, $\sigma_{Mg} = 0.07$ dex, $\sigma_{Si} = 0.04$ dex, and $\sigma_{Fe} = 0.02$ dex \citep[][]{Adibekyan-12c, Delgado-21}. A Monte Carlo approach was used to estimate the uncertainties in $Z$. 

It is important to note that we use present-day stellar abundances as a proxy for the primordial composition of protoplanetary disks. Some astrophysical processes, such as atomic diffusion, can alter a star’s surface composition as it evolves \citep[e.g.][]{Deal-18}, meaning the present-day abundances may differ from their primordial values. However, as we recently demonstrated in Adibekyan et al. (2024, A\&A, submitted), while the individual abundances of elements can change significantly over the evolution of FGK stars, the abundance ratios of rock-forming elements remain relatively constant. This consistency supports the validity of using present-day stellar abundances to infer the composition of protoplanetary disks.

As a measure of the metal fraction in the protoplanetary disk, $Z$ is independent of the disk mass. To calculate the total amount of metals ($Z_{\mathrm{total}}$) in the disks, one could multiply $Z$ by the protoplanetary disk mass. However, since these disks dissipated long ago, their masses cannot be directly determined. Despite significant progress in recent years toward understanding the relationship between protoplanetary disk masses and stellar properties, a robust quantitative expression for this relationship remains elusive. Both linear \citep[e.g.,][]{Andrews-13, Mohanty-13} and super-linear \citep[e.g.,][]{Ansdell-16, Pascucci-16} relationships between disk mass and stellar mass have been proposed, with substantial scatter around these trends \citep[e.g.,][]{Fiorellino-23, Guzman-diaz-23, Manara-23, Liu-24}.

Given the uncertainties in this relationship, we considered two approaches. Our primary approach assumes a linear relationship between protoplanetary disk mass and stellar mass, achieved by multiplying $Z$ by the stellar mass in units of solar mass. For the Solar System, this results in $Z_{\mathrm{total}} = 1.25$. The uncertainties in $Z_{\mathrm{total}}$ were determined by propagating the errors associated with $Z$ and stellar mass, while ignoring the unknown uncertainties related to the disk mass–stellar mass relationship.

Additionally, given the substantial dispersion ($\sigma_{\mathrm{disk}} \sim 0.6$–$0.9$ dex) observed in disk mass for a given stellar mass \citep[see discussion in][]{Manara-23}, we also adopted a conservative approach to estimate $Z^{*}_{\mathrm{total}}$ and its uncertainty. Since the dispersion is symmetric in the logarithmic scale, we calculated the total metal content as 
\[
Z^{*}_{\mathrm{total}} = \log(Z \times M_{\mathrm{star}}) \times \sigma_{\mathrm{disk}},
\] 
where \(\sigma_{\mathrm{disk}}\) represents the dispersion, assumed to be \(0.8\) dex. 

To account for the uncertainties in \(Z\) and \(M_{\mathrm{star}}\) for each target, as well as the dispersion \(\sigma_{\mathrm{disk}}\), we employed a Monte Carlo approach to generate a distribution for \(Z^{*}_{\mathrm{total}}\). Specifically, we drew \(10,000\) samples for \(Z\) and \(M_{\mathrm{star}}\) from Gaussian distributions parameterized by their respective uncertainties. Additionally, \(\sigma_{\mathrm{disk}}\) was sampled from a log-normal distribution centered at \(1\) with a standard deviation of \(0.8\) dex. The mean of the resulting \(Z^{*}_{\mathrm{total}}\) distribution was adopted as the best estimate, while its standard deviation was taken as the uncertainty. It is important to note that the uncertainties are primarily driven by \(\sigma_{\mathrm{disk}}\), with the contributions from the uncertainties in \(Z\) and \(M_{\mathrm{star}}\) being virtually negligible.

\section{Results} \label{sec:results}

Figure\, 1 shows that, overall, the host stars of Jupiters and super-Jupiters exhibit similar distributions of stellar mass and metallicity. The mean stellar mass of the hosts of Jupiters is 1.26$\pm$0.34 (0.01)\footnote{Values in parentheses represent the standard error of the mean.}, while for super-Jupiters, it is 1.28$\pm$0.48 (0.01). The mean [Fe/H] values for the two samples are 0.13$\pm$0.18 dex for Jupiters and 0.14$\pm$0.21 dex for super-Jupiters. A two-sample t-test (implemented via \texttt{scipy.stats.ttest\_ind}) was conducted to assess whether these mean values differ significantly, yielding p-values of 0.6 for stellar mass and 0.7 for metallicity, indicating no statistically significant differences.

Additionally, we performed Kolmogorov-Smirnov (KS) and Anderson-Darling (AD) tests to evaluate whether the distributions of stellar masses and metallicities for the two samples come from the same parent distribution. Both tests produced p-values greater than 0.25, further suggesting that the stellar masses and metallicities of super-Jupiters and Jupiters are not drawn from different parent populations.

These findings differ from earlier studies that suggested hosts of super-Jupiters are slightly less metal-rich \citep[e.g.][]{Santos-17a, Schlaufman-18}, particularly for more massive stars \citep{Adibekyan-19}. We tested our results using only the homogeneously determined metallicities from SWEET-Cat, and the results remained consistent. Therefore, the observed differences likely stem from the updated sample used in this work. Notably, the mean values obtained in both the current and previous studies are not significantly different.

The left panel of Figure 2 presents the relation between $Z$ and planetary mass. Super-Jupiters show a slightly higher average $Z$ than Jupiters, with values of 1.79$\pm$0.58 (0.01) compared to 1.63$\pm$0.58 (0.01). To assess the statistical significance of this difference, we conducted a two-sample t-test, which resulted in a p-value of 0.035, just below the conventional threshold for statistical significance. We also applied KS and AD tests to investigate whether the $Z_{\mathrm{total}}$ distributions of the two groups come from the same parent population. The tests yielded p-values of 0.031 and 0.037, respectively, indicating the possibility of a distinction between the distributions.

\begin{figure}[ht!]
\includegraphics[width = 1\textwidth]{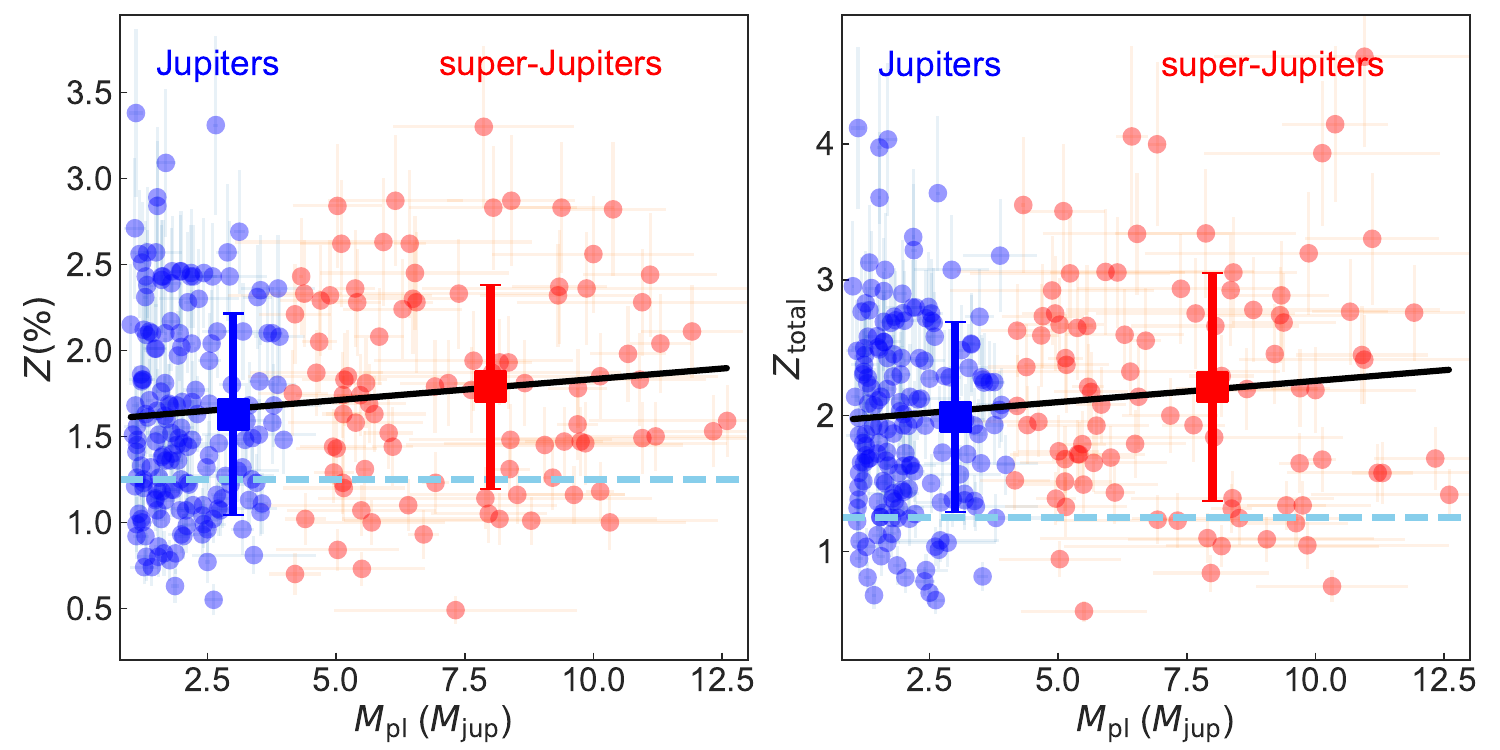}
\caption{Metal mass fraction (left panel) and total metal content (right panel) of the protoplanetary disk as a function of planetary mass. Blue and red circles represent Jupiters and super-Jupiters, respectively, while large blue and red squares show the mean values and standard deviations for each group. Black lines show the linear regression results, and the sky-blue dashed lines indicate the $Z$ and $Z_{\mathrm{total}}$ values for the Solar System. \label{fig:2}}
\end{figure} 

In the right panel of Figure 2 we show how $Z_{\mathrm{total}}$ varies with planetary mass. We observe that only 13\% (12 out of 93) of super-Jupiters and 15\% (25 out of 172) of Jupiters have $Z_{\mathrm{total}}$ values below the proto-solar disk level. The average $Z_{\mathrm{total}}$ for super-Jupiters is slightly higher than that for Jupiters: 2.21$\pm$0.84 (0.01) compared to 1.99$\pm$0.69 (0.01). A two-sample t-test gives a p-value of 0.023, indicating a statistically significant difference. The KS and AD tests further suggest that the $Z_{\mathrm{total}}$ distributions of super-Jupiters and Jupiters differ, returning p-values of 0.034 and 0.042, respectively.

To further explore the relationship between planetary mass and the total amount of metals available in the protoplanetary disk, we applied Ordinary Least Squares (OLS) regression using \texttt{statsmodels.regression.linear\_model.OLS}). The resulting fit is shown as a solid line in Figure 2. A slight correlation is observed (slope = 0.031$\pm$0.016), with a p-value of 0.048 from the F-statistic, indicating a potentially statistically significant relationship.

To fully account for the uncertainties in both planetary mass and $Z_{\mathrm{total}}$, we performed a MC test. Assuming Gaussian distributions for the uncertainties in these parameters, with means centered on their respective values, we randomly drew values for $Z_{\mathrm{total}}$ and $M_{\mathrm{pl}}$ for the targets. We then split the sample into Jupiters and super-Jupiters based on the criteria outlined in the manuscript, using a boundary at 4 $M_{\mathrm{Jup}}$. Subsequently, we performed KS and AD tests, as well as a two-sample t-test, to compare the mean values of $Z_{\mathrm{total}}$. We conducted 10,000 such realizations. In less than 50\% of cases, the p-values from the aforementioned tests were below the standard significance threshold of 0.05. Therefore, this test does not support the hypothesis that the two samples have statistically distinct distributions or mean values of $Z_{\mathrm{total}}$. However, in only 1 out of 10,000 realizations, the mean value of $Z_{\mathrm{total}}$ for super-Jupiters was smaller than that for Jupiters.

Additionally, we repeated this entire test, but instead of assuming a linear relationship between disk mass and stellar mass, we assumed a superlinear relationship with a power of 1.5. In only 4 realizations was $Z_{\mathrm{total}}$ for super-Jupiters smaller than that for Jupiters. 

Figure 3 shows $Z^{*}_{\mathrm{total}}$ as a function of planetary mass, also displaying the mean values and standard deviations for the two groups. We repeated the same MC test, but for $Z^{*}_{\mathrm{total}}$. In over $94\%$ of realizations, the p-values for the KS test, AD test, and two-sample t-test were above the typical significance threshold of $0.05$. Additionally, in about $64\%$ of realizations, the mean values of $Z^{*}_{\mathrm{total}}$ for super-Jupiters were higher than those for Jupiters. These results, suggesting similar distributions for the two samples, are expected given the sensitivity of the test to the uncertainties and the large uncertainties associated with $Z^{*}_{\mathrm{total}}$.

\begin{figure}[ht!]
\includegraphics[width = 0.7\textwidth]{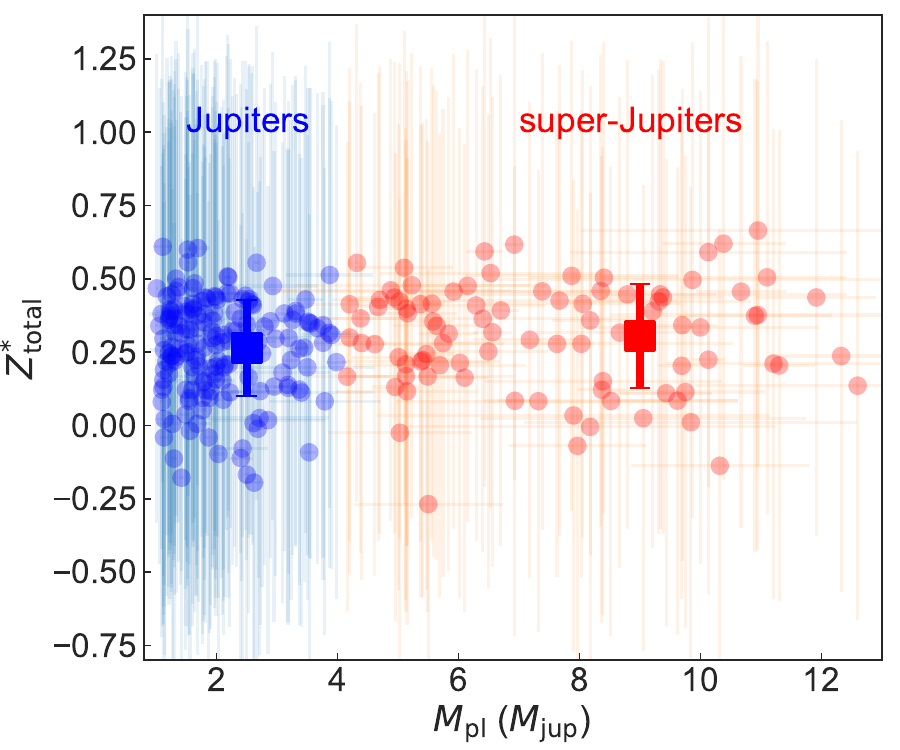}
\caption{Total metal content ($Z^{*}_{\mathrm{total}}$) of the protoplanetary disk as a function of planetary mass. Blue and red circles represent Jupiters and super-Jupiters, respectively, while large blue and red squares show the mean values and standard deviations for each group. \label{fig:3}}
\end{figure} 

We note that there are 10 systems with planets present in both the Jupiter and super-Jupiter groups. These systems were not excluded when performing the statistical tests. However, excluding them does not affect the outcomes of the tests.

\section{Discussion and Conclusion} \label{sec:conclusion}

Iron abundance can be used as a proxy for overall metallicity in stars with approximately solar metallicity. However, at lower [Fe/H], stars tend to be enriched (relative to iron) in elements such as carbon, oxygen, magnesium, and silicon \citep[e.g.][]{Adibekyan-12, Bensby-14, Delgado-21},  all of which are crucial for the formation of massive planets. Therefore, especially at low iron abundances, the total mass fraction of heavy elements should be used to estimate the metal content \citep[e.g.][]{Gonzalez-09, Adibekyan-12b}. 

The observational evidence that Jupiter-mass planets tend to form around stars with high metallicity (typically [Fe/H] has been used as a proxy as a proxy) supports their formation via CA \citep[e.g.][]{Ida-04, Mordasini-12}. Our results show that super-massive planets typically form in disks with even slightly higher metal content, mostly exceeding that of the proto-solar disk. This suggests that the formation of super-massive planets is feasible through the CA mechanism.

The observational evidence that Jupiter-mass planets tend to form around stars with high metallicity (typically [Fe/H] has been used as a proxy as a proxy) supports their formation via CA \citep[e.g.][]{Ida-04, Mordasini-12}. Although some of the statistical tests we performed did not support the hypothesis that the distributions or mean values of $Z_{\mathrm{total}}$ and $Z^{*}_{\mathrm{total}}$ for Jupiters and super-Jupiters are statistically different, the results consistently demonstrate that super-massive planets form in disks with at least as much metal content as those that host Jupiters, and often with even slightly higher metallicity, generally exceeding that of the proto-solar disk. This suggests that the formation of super-massive planets is feasible through the CA mechanism.

Interestingly, we also observed 12 super-Jupiters  with $Z_{\mathrm{total}}$ values lower than the proto-solar value, and we did not identify any clear pattern distinguishing these targets from the other super-Jupiters. The formation of such massive planets in sub-solar metallicity protoplanetary disks is challenging to explain solely through CA \citep{Mordasini-12}, but could potentially be accounted for by GI \citep{Matsukoba-23}.

In summary, our findings provide strong evidence that super-massive planets can form through CA. Nevertheless, further theoretical modeling is needed to understand the formation of such planets in metal-poor disks \citep[see e.g.][]{Matsukoba-23}, particularly around massive stars, and we encourage such studies to confirm our observational conclusions. 

The ongoing Gaia mission \citep{Gaia-16} and the upcoming PLATO mission \citep[PLAnetary Transits and Oscillations of Stars,][]{Rauer-24} are expected to detect a large number of massive planets. Notably, PLATO will provide precise measurements of host star masses. Follow-up spectroscopic observations of these planets to determine their masses (for transiting planets), along with homogeneous abundance measurements of the host stars, would significantly enhance the robustness of our results and represent a valuable improvement to this work.

\backmatter

\bmhead{Acknowledgements}

V.A. was supported by Funda\c{c}\~ao para a Ci\^encia e Tecnologia through national funds and by FEDER through COMPETE2020 - Programa Operacional Competitividade e Internacionalização by these grants: UIDB/04434/2020; UIDP/04434/2020; 2022.06962.PTDC.

This research has made use of the NASA Exoplanet Archive, which is operated by the California Institute of Technology, under contract with the National Aeronautics and Space Administration under the Exoplanet Exploration Program.
The research shown here acknowledges use of the Hypatia Catalog Database, an online compilation of stellar abundance data as described in \citet{Hinkel-14}, which was supported by NASA's Nexus for Exoplanet System Science (NExSS) research coordination network and the Vanderbilt Initiative in Data-Intensive Astrophysics (VIDA).

 In this work we used the Python language and several scientific packages: {SciPy \citep{Virtanen-20}, Matplotlib \citep{Hunter-07}, Statsmodels \citep{Seabold-10}, Pandas \citep{mckinney2010data}, Seaborn \citep{Waskom2021}}.

\bibliography{sn-article}

\begin{thebibliography}{42}
\providecommand{\natexlab}[1]{#1}
\providecommand{\url}[1]{{#1}}
\providecommand{\urlprefix}{URL }
\providecommand{\doi}[1]{\url{https://doi.org/#1}}
\providecommand{\eprint}[2][]{\url{#2}}
 \bibcommenthead

\bibitem[{{Adibekyan}(2019)}]{Adibekyan-19}
{Adibekyan} V (2019) {Heavy Metal Rules. I. Exoplanet Incidence and
  Metallicity}. Geosciences 9(3):105. \doi{10.3390/geosciences9030105},
  {\href{https://arxiv.org/abs/1902.04493}{{arXiv:1902.04493}}} {[astro-ph.EP]}

\bibitem[{{Adibekyan} et~al(2012{\natexlab{a}}){Adibekyan}, {Delgado Mena},
  {Sousa}, {Santos}, {Israelian}, {Gonz{\'a}lez Hern{\'a}ndez}, {Mayor}, and
  {Hakobyan}}]{Adibekyan-12}
{Adibekyan} VZ, {Delgado Mena} E, {Sousa} SG, et~al (2012{\natexlab{a}})
  {Exploring the {\ensuremath{\alpha}}-enhancement of metal-poor planet-hosting
  stars. The Kepler and HARPS samples}. \aap 547:A36.
  \doi{10.1051/0004-6361/201220167},
  {\href{https://arxiv.org/abs/1209.6272}{{arXiv:1209.6272}}} {[astro-ph.EP]}

\bibitem[{{Adibekyan} et~al(2012{\natexlab{b}}){Adibekyan}, {Santos}, {Sousa},
  {Israelian}, {Delgado Mena}, {Gonz{\'a}lez Hern{\'a}ndez}, {Mayor}, {Lovis},
  and {Udry}}]{Adibekyan-12b}
{Adibekyan} VZ, {Santos} NC, {Sousa} SG, et~al (2012{\natexlab{b}})
  {Overabundance of {\ensuremath{\alpha}}-elements in exoplanet-hosting stars}.
  \aap 543:A89. \doi{10.1051/0004-6361/201219564},
  {\href{https://arxiv.org/abs/1205.6670}{{arXiv:1205.6670}}} {[astro-ph.EP]}

\bibitem[{{Adibekyan} et~al(2012{\natexlab{c}}){Adibekyan}, {Sousa}, {Santos},
  {Delgado Mena}, {Gonz{\'a}lez Hern{\'a}ndez}, {Israelian}, {Mayor}, and
  {Khachatryan}}]{Adibekyan-12c}
{Adibekyan} VZ, {Sousa} SG, {Santos} NC, et~al (2012{\natexlab{c}}) {Chemical
  abundances of 1111 FGK stars from the HARPS GTO planet search program.
  Galactic stellar populations and planets}. \aap 545:A32.
  \doi{10.1051/0004-6361/201219401},
  {\href{https://arxiv.org/abs/1207.2388}{{arXiv:1207.2388}}} {[astro-ph.EP]}

\bibitem[{{Adibekyan} et~al(2013){Adibekyan}, {Figueira}, {Santos}, {Mortier},
  {Mordasini}, {Delgado Mena}, {Sousa}, {Correia}, {Israelian}, and
  {Oshagh}}]{Adibekyan-13}
{Adibekyan} VZ, {Figueira} P, {Santos} NC, et~al (2013) {Orbital and physical
  properties of planets and their hosts: new insights on planet formation and
  evolution}. \aap 560:A51. \doi{10.1051/0004-6361/201322551},
  {\href{https://arxiv.org/abs/1311.2417}{{arXiv:1311.2417}}} {[astro-ph.EP]}

\bibitem[{{Andrews} et~al(2013){Andrews}, {Rosenfeld}, {Kraus}, and
  {Wilner}}]{Andrews-13}
{Andrews} SM, {Rosenfeld} KA, {Kraus} AL, et~al (2013) {The Mass Dependence
  between Protoplanetary Disks and their Stellar Hosts}. \apj 771(2):129.
  \doi{10.1088/0004-637X/771/2/129},
  {\href{https://arxiv.org/abs/1305.5262}{{arXiv:1305.5262}}} {[astro-ph.SR]}

\bibitem[{{Ansdell} et~al(2016){Ansdell}, {Williams}, {van der Marel},
  {Carpenter}, {Guidi}, {Hogerheijde}, {Mathews}, {Manara}, {Miotello},
  {Natta}, {Oliveira}, {Tazzari}, {Testi}, {van Dishoeck}, and {van
  Terwisga}}]{Ansdell-16}
{Ansdell} M, {Williams} JP, {van der Marel} N, et~al (2016) {ALMA Survey of
  Lupus Protoplanetary Disks. I. Dust and Gas Masses}. \apj 828(1):46.
  \doi{10.3847/0004-637X/828/1/46},
  {\href{https://arxiv.org/abs/1604.05719}{{arXiv:1604.05719}}} {[astro-ph.EP]}

\bibitem[{{Asplund} et~al(2021){Asplund}, {Amarsi}, and
  {Grevesse}}]{Asplund-21}
{Asplund} M, {Amarsi} AM, {Grevesse} N (2021) {The chemical make-up of the Sun:
  A 2020 vision}. \aap 653:A141. \doi{10.1051/0004-6361/202140445},
  {\href{https://arxiv.org/abs/2105.01661}{{arXiv:2105.01661}}} {[astro-ph.SR]}

\bibitem[{{Bensby} et~al(2014){Bensby}, {Feltzing}, and {Oey}}]{Bensby-14}
{Bensby} T, {Feltzing} S, {Oey} MS (2014) {Exploring the Milky Way stellar
  disk. A detailed elemental abundance study of 714 F and G dwarf stars in the
  solar neighbourhood}. \aap 562:A71. \doi{10.1051/0004-6361/201322631},
  {\href{https://arxiv.org/abs/1309.2631}{{arXiv:1309.2631}}} {[astro-ph.GA]}

\bibitem[{{Boss}(1997)}]{Boss-97}
{Boss} AP (1997) {Giant planet formation by gravitational instability.} Science
  276:1836--1839. \doi{10.1126/science.276.5320.1836}

\bibitem[{{Deal} et~al(2018){Deal}, {Alecian}, {Lebreton}, {Goupil}, {Marques},
  {LeBlanc}, {Morel}, and {Pichon}}]{Deal-18}
{Deal} M, {Alecian} G, {Lebreton} Y, et~al (2018) {Impacts of radiative
  accelerations on solar-like oscillating main-sequence stars}. \aap 618:A10.
  \doi{10.1051/0004-6361/201833361},
  {\href{https://arxiv.org/abs/1806.10533}{{arXiv:1806.10533}}} {[astro-ph.SR]}

\bibitem[{{Delgado Mena} et~al(2021){Delgado Mena}, {Adibekyan}, {Santos},
  {Tsantaki}, {Gonz{\'a}lez Hern{\'a}ndez}, {Sousa}, and {Bertr{\'a}n de
  Lis}}]{Delgado-21}
{Delgado Mena} E, {Adibekyan} V, {Santos} NC, et~al (2021) {Chemical abundances
  of 1111 FGK stars from the HARPS GTO planet search program. IV. Carbon and
  C/O ratios for Galactic stellar populations and planet hosts}. \aap 655:A99.
  \doi{10.1051/0004-6361/202141588},
  {\href{https://arxiv.org/abs/2109.04844}{{arXiv:2109.04844}}} {[astro-ph.SR]}

\bibitem[{{Fiorellino} et~al(2023){Fiorellino}, {Tychoniec}, {Cruz-S{\'a}enz de
  Miera}, {Antoniucci}, {K{\'o}sp{\'a}l}, {Manara}, {Nisini}, and
  {Rosotti}}]{Fiorellino-23}
{Fiorellino} E, {Tychoniec} {\L}, {Cruz-S{\'a}enz de Miera} F, et~al (2023)
  {The Mass Accretion Rate and Stellar Properties in Class I Protostars}. \apj
  944(2):135. \doi{10.3847/1538-4357/aca320},
  {\href{https://arxiv.org/abs/2211.07653}{{arXiv:2211.07653}}} {[astro-ph.SR]}

\bibitem[{{Fischer} and {Valenti}(2005)}]{Fischer-05}
{Fischer} DA, {Valenti} J (2005) {The Planet-Metallicity Correlation}. \apj
  622(2):1102--1117. \doi{10.1086/428383}

\bibitem[{{Gaia Collaboration} et~al(2016){Gaia Collaboration}, {Prusti}, {de
  Bruijne}, {Brown}, {Vallenari}, {Babusiaux}, {Bailer-Jones}, {Bastian},
  {Biermann}, {Evans}, {Eyer}, {Jansen}, {Jordi}, {Klioner}, {Lammers},
  {Lindegren}, {Luri}, {Mignard}, {Milligan}, {Panem}, {Poinsignon},
  {Pourbaix}, {Randich}, {Sarri}, {Sartoretti}, {Siddiqui}, {Soubiran},
  {Valette}, {van Leeuwen}, {Walton}, {Aerts}, {Arenou}, {Cropper}, {Drimmel},
  {H{\o}g}, {Katz}, {Lattanzi}, {O'Mullane}, {Grebel}, {Holland}, {Huc},
  {Passot}, {Bramante}, {Cacciari}, {Casta{\~n}eda}, {Chaoul}, {Cheek}, {De
  Angeli}, {Fabricius}, {Guerra}, {Hern{\'a}ndez}, {Jean-Antoine-Piccolo},
  {Masana}, {Messineo}, {Mowlavi}, {Nienartowicz}, {Ord{\'o}{\~n}ez-Blanco},
  {Panuzzo}, {Portell}, {Richards}, {Riello}, {Seabroke}, {Tanga},
  {Th{\'e}venin}, {Torra}, {Els}, {Gracia-Abril}, {Comoretto},
  {Garcia-Reinaldos}, {Lock}, {Mercier}, {Altmann}, {Andrae}, {Astraatmadja},
  {Bellas-Velidis}, {Benson}, {Berthier}, {Blomme}, {Busso}, {Carry},
  {Cellino}, {Clementini}, {Cowell}, {Creevey}, {Cuypers}, {Davidson}, {De
  Ridder}, {de Torres}, {Delchambre}, {Dell'Oro}, {Ducourant}, {Fr{\'e}mat},
  {Garc{\'\i}a-Torres}, {Gosset}, {Halbwachs}, {Hambly}, {Harrison}, {Hauser},
  {Hestroffer}, {Hodgkin}, {Huckle}, {Hutton}, {Jasniewicz}, {Jordan},
  {Kontizas}, {Korn}, {Lanzafame}, {Manteiga}, {Moitinho}, {Muinonen},
  {Osinde}, {Pancino}, {Pauwels}, {Petit}, {Recio-Blanco}, {Robin}, {Sarro},
  {Siopis}, {Smith}, {Smith}, {Sozzetti}, {Thuillot}, {van Reeven}, {Viala},
  {Abbas}, {Abreu Aramburu}, {Accart}, {Aguado}, {Allan}, {Allasia},
  {Altavilla}, {{\'A}lvarez}, {Alves}, {Anderson}, {Andrei}, {Anglada Varela},
  {Antiche}, {Antoja}, {Ant{\'o}n}, {Arcay}, {Atzei}, {Ayache}, {Bach},
  {Baker}, {Balaguer-N{\'u}{\~n}ez}, {Barache}, {Barata}, {Barbier}, {Barblan},
  {Baroni}, {Barrado y Navascu{\'e}s}, {Barros}, {Barstow}, {Becciani},
  {Bellazzini}, {Bellei}, {Bello Garc{\'\i}a}, {Belokurov}, {Bendjoya},
  {Berihuete}, {Bianchi}, {Bienaym{\'e}}, {Billebaud}, {Blagorodnova},
  {Blanco-Cuaresma}, {Boch}, {Bombrun}, {Borrachero}, {Bouquillon}, {Bourda},
  {Bouy}, {Bragaglia}, {Breddels}, {Brouillet}, {Br{\"u}semeister},
  {Bucciarelli}, {Budnik}, {Burgess}, {Burgon}, {Burlacu}, {Busonero}, {Buzzi},
  {Caffau}, {Cambras}, {Campbell}, {Cancelliere}, {Cantat-Gaudin}, {Carlucci},
  {Carrasco}, {Castellani}, {Charlot}, {Charnas}, {Charvet}, {Chassat},
  {Chiavassa}, {Clotet}, {Cocozza}, {Collins}, {Collins}, {Costigan}, {Crifo},
  {Cross}, {Crosta}, {Crowley}, {Dafonte}, {Damerdji}, {Dapergolas}, {David},
  {David}, {De Cat}, {de Felice}, {de Laverny}, {De Luise}, {De March}, {de
  Martino}, {de Souza}, {Debosscher}, {del Pozo}, {Delbo}, {Delgado},
  {Delgado}, {di Marco}, {Di Matteo}, {Diakite}, {Distefano}, {Dolding}, {Dos
  Anjos}, {Drazinos}, {Dur{\'a}n}, {Dzigan}, {Ecale}, {Edvardsson}, {Enke},
  {Erdmann}, {Escolar}, {Espina}, {Evans}, {Eynard Bontemps}, {Fabre},
  {Fabrizio}, {Faigler}, {Falc{\~a}o}, {Farr{\`a}s Casas}, {Faye}, {Federici},
  {Fedorets}, {Fern{\'a}ndez-Hern{\'a}ndez}, {Fernique}, {Fienga}, {Figueras},
  {Filippi}, {Findeisen}, {Fonti}, {Fouesneau}, {Fraile}, {Fraser}, {Fuchs},
  {Furnell}, {Gai}, {Galleti}, {Galluccio}, {Garabato}, {Garc{\'\i}a-Sedano},
  {Gar{\'e}}, {Garofalo}, {Garralda}, {Gavras}, {Gerssen}, {Geyer}, {Gilmore},
  {Girona}, {Giuffrida}, {Gomes}, {Gonz{\'a}lez-Marcos},
  {Gonz{\'a}lez-N{\'u}{\~n}ez}, {Gonz{\'a}lez-Vidal}, {Granvik}, {Guerrier},
  {Guillout}, {Guiraud}, {G{\'u}rpide}, {Guti{\'e}rrez-S{\'a}nchez}, {Guy},
  {Haigron}, {Hatzidimitriou}, {Haywood}, {Heiter}, {Helmi}, {Hobbs},
  {Hofmann}, {Holl}, {Holland}, {Hunt}, {Hypki}, {Icardi}, {Irwin}, {Jevardat
  de Fombelle}, {Jofr{\'e}}, {Jonker}, {Jorissen}, {Julbe}, {Karampelas},
  {Kochoska}, {Kohley}, {Kolenberg}, {Kontizas}, {Koposov}, {Kordopatis},
  {Koubsky}, {Kowalczyk}, {Krone-Martins}, {Kudryashova}, {Kull}, {Bachchan},
  {Lacoste-Seris}, {Lanza}, {Lavigne}, {Le Poncin-Lafitte}, {Lebreton},
  {Lebzelter}, {Leccia}, {Leclerc}, {Lecoeur-Taibi}, {Lemaitre}, {Lenhardt},
  {Leroux}, {Liao}, {Licata}, {Lindstr{\o}m}, {Lister}, {Livanou}, {Lobel},
  {L{\"o}ffler}, {L{\'o}pez}, {Lopez-Lozano}, {Lorenz}, {Loureiro},
  {MacDonald}, {Magalh{\~a}es Fernandes}, {Managau}, {Mann}, {Mantelet},
  {Marchal}, {Marchant}, {Marconi}, {Marie}, {Marinoni}, {Marrese},
  {Marschalk{\'o}}, {Marshall}, {Mart{\'\i}n-Fleitas}, {Martino}, {Mary},
  {Matijevi{\v{c}}}, {Mazeh}, {McMillan}, {Messina}, {Mestre}, {Michalik},
  {Millar}, {Miranda}, {Molina}, {Molinaro}, {Molinaro}, {Moln{\'a}r},
  {Moniez}, {Montegriffo}, {Monteiro}, {Mor}, {Mora}, {Morbidelli}, {Morel},
  {Morgenthaler}, {Morley}, {Morris}, {Mulone}, {Muraveva}, {Musella},
  {Narbonne}, {Nelemans}, {Nicastro}, {Noval}, {Ord{\'e}novic},
  {Ordieres-Mer{\'e}}, {Osborne}, {Pagani}, {Pagano}, {Pailler}, {Palacin},
  {Palaversa}, {Parsons}, {Paulsen}, {Pecoraro}, {Pedrosa}, {Pentik{\"a}inen},
  {Pereira}, {Pichon}, {Piersimoni}, {Pineau}, {Plachy}, {Plum}, {Poujoulet},
  {Pr{\v{s}}a}, {Pulone}, {Ragaini}, {Rago}, {Rambaux}, {Ramos-Lerate},
  {Ranalli}, {Rauw}, {Read}, {Regibo}, {Renk}, {Reyl{\'e}}, {Ribeiro},
  {Rimoldini}, {Ripepi}, {Riva}, {Rixon}, {Roelens}, {Romero-G{\'o}mez},
  {Rowell}, {Royer}, {Rudolph}, {Ruiz-Dern}, {Sadowski}, {Sagrist{\`a}
  Sell{\'e}s}, {Sahlmann}, {Salgado}, {Salguero}, {Sarasso}, {Savietto},
  {Schnorhk}, {Schultheis}, {Sciacca}, {Segol}, {Segovia}, {Segransan},
  {Serpell}, {Shih}, {Smareglia}, {Smart}, {Smith}, {Solano}, {Solitro},
  {Sordo}, {Soria Nieto}, {Souchay}, {Spagna}, {Spoto}, {Stampa}, {Steele},
  {Steidelm{\"u}ller}, {Stephenson}, {Stoev}, {Suess}, {S{\"u}veges}, {Surdej},
  {Szabados}, {Szegedi-Elek}, {Tapiador}, {Taris}, {Tauran}, {Taylor},
  {Teixeira}, {Terrett}, {Tingley}, {Trager}, {Turon}, {Ulla}, {Utrilla},
  {Valentini}, {van Elteren}, {Van Hemelryck}, {van Leeuwen}, {Varadi},
  {Vecchiato}, {Veljanoski}, {Via}, {Vicente}, {Vogt}, {Voss}, {Votruba},
  {Voutsinas}, {Walmsley}, {Weiler}, {Weingrill}, {Werner}, {Wevers},
  {Whitehead}, {Wyrzykowski}, {Yoldas}, {{\v{Z}}erjal}, {Zucker}, {Zurbach},
  {Zwitter}, {Alecu}, {Allen}, {Allende Prieto}, {Amorim},
  {Anglada-Escud{\'e}}, {Arsenijevic}, {Azaz}, {Balm}, {Beck}, {Bernstein},
  {Bigot}, {Bijaoui}, {Blasco}, {Bonfigli}, {Bono}, {Boudreault}, {Bressan},
  {Brown}, {Brunet}, {Bunclark}, {Buonanno}, {Butkevich}, {Carret}, {Carrion},
  {Chemin}, {Ch{\'e}reau}, {Corcione}, {Darmigny}, {de Boer}, {de Teodoro}, {de
  Zeeuw}, {Delle Luche}, {Domingues}, {Dubath}, {Fodor}, {Fr{\'e}zouls},
  {Fries}, {Fustes}, {Fyfe}, {Gallardo}, {Gallegos}, {Gardiol}, {Gebran},
  {Gomboc}, {G{\'o}mez}, {Grux}, {Gueguen}, {Heyrovsky}, {Hoar}, {Iannicola},
  {Isasi Parache}, {Janotto}, {Joliet}, {Jonckheere}, {Keil}, {Kim},
  {Klagyivik}, {Klar}, {Knude}, {Kochukhov}, {Kolka}, {Kos}, {Kutka}, {Lainey},
  {LeBouquin}, {Liu}, {Loreggia}, {Makarov}, {Marseille}, {Martayan},
  {Martinez-Rubi}, {Massart}, {Meynadier}, {Mignot}, {Munari}, {Nguyen},
  {Nordlander}, {Ocvirk}, {O'Flaherty}, {Olias Sanz}, {Ortiz}, {Osorio},
  {Oszkiewicz}, {Ouzounis}, {Palmer}, {Park}, {Pasquato}, {Peltzer}, {Peralta},
  {P{\'e}turaud}, {Pieniluoma}, {Pigozzi}, {Poels}, {Prat}, {Prod'homme},
  {Raison}, {Rebordao}, {Risquez}, {Rocca-Volmerange}, {Rosen}, {Ruiz-Fuertes},
  {Russo}, {Sembay}, {Serraller Vizcaino}, {Short}, {Siebert}, {Silva},
  {Sinachopoulos}, {Slezak}, {Soffel}, {Sosnowska}, {Strai{\v{z}}ys}, {ter
  Linden}, {Terrell}, {Theil}, {Tiede}, {Troisi}, {Tsalmantza}, {Tur},
  {Vaccari}, {Vachier}, {Valles}, {Van Hamme}, {Veltz}, {Virtanen}, {Wallut},
  {Wichmann}, {Wilkinson}, {Ziaeepour}, and {Zschocke}}]{Gaia-16}
{Gaia Collaboration}, {Prusti} T, {de Bruijne} JHJ, et~al (2016) {The Gaia
  mission}. \aap 595:A1. \doi{10.1051/0004-6361/201629272},
  {\href{https://arxiv.org/abs/1609.04153}{{arXiv:1609.04153}}} {[astro-ph.IM]}

\bibitem[{{Gonzalez}(2009)}]{Gonzalez-09}
{Gonzalez} G (2009) {Stars with planets and the thick disc}. \mnras
  399(1):L103--L107. \doi{10.1111/j.1745-3933.2009.00734.x}

\bibitem[{{Guzm{\'a}n-D{\'\i}az} et~al(2023){Guzm{\'a}n-D{\'\i}az},
  {Montesinos}, {Mendigut{\'\i}a}, {Kama}, {Meeus}, {Vioque}, {Oudmaijer}, and
  {Villaver}}]{Guzman-diaz-23}
{Guzm{\'a}n-D{\'\i}az} J, {Montesinos} B, {Mendigut{\'\i}a} I, et~al (2023)
  {Relation between metallicities and spectral energy distributions of Herbig
  Ae/Be stars. A potential link with planet formation}. \aap 671:A140.
  \doi{10.1051/0004-6361/202245427},
  {\href{https://arxiv.org/abs/2212.14022}{{arXiv:2212.14022}}} {[astro-ph.SR]}

\bibitem[{{Hinkel} et~al(2014){Hinkel}, {Timmes}, {Young}, {Pagano}, and
  {Turnbull}}]{Hinkel-14}
{Hinkel} NR, {Timmes} FX, {Young} PA, et~al (2014) {Stellar Abundances in the
  Solar Neighborhood: The Hypatia Catalog}. \aj 148(3):54.
  \doi{10.1088/0004-6256/148/3/54},
  {\href{https://arxiv.org/abs/1405.6719}{{arXiv:1405.6719}}} {[astro-ph.SR]}

\bibitem[{Hunter(2007)}]{Hunter-07}
Hunter JD (2007) Matplotlib: A 2d graphics environment. Computing in Science \&
  Engineering 9(3):90--95. \doi{10.1109/MCSE.2007.55}

\bibitem[{{Ida} and {Lin}(2004)}]{Ida-04}
{Ida} S, {Lin} DNC (2004) {Toward a Deterministic Model of Planetary Formation.
  I. A Desert in the Mass and Semimajor Axis Distributions of Extrasolar
  Planets}. \apj 604(1):388--413. \doi{10.1086/381724},
  {\href{https://arxiv.org/abs/astro-ph/0312144}{{arXiv:astro-ph/0312144}}}
  {[astro-ph]}

\bibitem[{{Liu} et~al(2024){Liu}, {Roussel}, {Linz}, {Fang}, {Wolf},
  {Kirchschlager}, {Henning}, {Yang}, {Du}, {Flock}, and {Wang}}]{Liu-24}
{Liu} Y, {Roussel} H, {Linz} H, et~al (2024) {Dust mass in protoplanetary disks
  with porous dust opacities}. arXiv e-prints arXiv:2411.00277.
  {\href{https://arxiv.org/abs/2411.00277}{{arXiv:2411.00277}}} {[astro-ph.EP]}

\bibitem[{{Maldonado} et~al(2019){Maldonado}, {Villaver}, {Eiroa}, and
  {Micela}}]{Maldonado-19}
{Maldonado} J, {Villaver} E, {Eiroa} C, et~al (2019) {Connecting substellar and
  stellar formation: the role of the host star's metallicity}. \aap 624:A94.
  \doi{10.1051/0004-6361/201833827},
  {\href{https://arxiv.org/abs/1903.01141}{{arXiv:1903.01141}}} {[astro-ph.SR]}

\bibitem[{{Manara} et~al(2023){Manara}, {Ansdell}, {Rosotti}, {Hughes},
  {Armitage}, {Lodato}, and {Williams}}]{Manara-23}
{Manara} CF, {Ansdell} M, {Rosotti} GP, et~al (2023) {Demographics of Young
  Stars and their Protoplanetary Disks: Lessons Learned on Disk Evolution and
  its Connection to Planet Formation}. In: {Inutsuka} S, {Aikawa} Y, {Muto} T,
  et~al (eds) Protostars and Planets VII, p 539,
  \doi{10.48550/arXiv.2203.09930}, \eprint{2203.09930}

\bibitem[{{Matsukoba} et~al(2023){Matsukoba}, {Vorobyov}, {Hosokawa}, and
  {Guedel}}]{Matsukoba-23}
{Matsukoba} R, {Vorobyov} EI, {Hosokawa} T, et~al (2023) {Formation of a
  wide-orbit giant planet in a gravitationally unstable subsolar-metallicity
  protoplanetary disc}. \mnras 526(3):3933--3943. \doi{10.1093/mnras/stad3003},
  {\href{https://arxiv.org/abs/2307.13722}{{arXiv:2307.13722}}} {[astro-ph.EP]}

\bibitem[{McKinney et~al(2010)}]{mckinney2010data}
McKinney W, et~al (2010) Data structures for statistical computing in python.
  In: Proceedings of the 9th Python in Science Conference, Austin, TX, pp
  51--56

\bibitem[{{Mohanty} et~al(2013){Mohanty}, {Greaves}, {Mortlock}, {Pascucci},
  {Scholz}, {Thompson}, {Apai}, {Lodato}, and {Looper}}]{Mohanty-13}
{Mohanty} S, {Greaves} J, {Mortlock} D, et~al (2013) {Protoplanetary Disk
  Masses from Stars to Brown Dwarfs}. \apj 773(2):168.
  \doi{10.1088/0004-637X/773/2/168},
  {\href{https://arxiv.org/abs/1305.6896}{{arXiv:1305.6896}}} {[astro-ph.SR]}

\bibitem[{{Mordasini} et~al(2012){Mordasini}, {Alibert}, {Benz}, {Klahr}, and
  {Henning}}]{Mordasini-12}
{Mordasini} C, {Alibert} Y, {Benz} W, et~al (2012) {Extrasolar planet
  population synthesis . IV. Correlations with disk metallicity, mass, and
  lifetime}. \aap 541:A97. \doi{10.1051/0004-6361/201117350},
  {\href{https://arxiv.org/abs/1201.1036}{{arXiv:1201.1036}}} {[astro-ph.EP]}

\bibitem[{{Narang} et~al(2018){Narang}, {Manoj}, {Furlan}, {Mordasini},
  {Henning}, {Mathew}, {Banyal}, and {Sivarani}}]{Narang-18}
{Narang} M, {Manoj} P, {Furlan} E, et~al (2018) {Properties and Occurrence
  Rates for Kepler Exoplanet Candidates as a Function of Host Star Metallicity
  from the DR25 Catalog}. \aj 156(5):221. \doi{10.3847/1538-3881/aae391},
  {\href{https://arxiv.org/abs/1809.08385}{{arXiv:1809.08385}}} {[astro-ph.EP]}

\bibitem[{{Pascucci} et~al(2016){Pascucci}, {Testi}, {Herczeg}, {Long},
  {Manara}, {Hendler}, {Mulders}, {Krijt}, {Ciesla}, {Henning}, {Mohanty},
  {Drabek-Maunder}, {Apai}, {Sz{\H{u}}cs}, {Sacco}, and
  {Olofsson}}]{Pascucci-16}
{Pascucci} I, {Testi} L, {Herczeg} GJ, et~al (2016) {A Steeper than Linear Disk
  Mass-Stellar Mass Scaling Relation}. \apj 831(2):125.
  \doi{10.3847/0004-637X/831/2/125},
  {\href{https://arxiv.org/abs/1608.03621}{{arXiv:1608.03621}}} {[astro-ph.EP]}

\bibitem[{{Pollack} et~al(1996){Pollack}, {Hubickyj}, {Bodenheimer},
  {Lissauer}, {Podolak}, and {Greenzweig}}]{Pollack-96}
{Pollack} JB, {Hubickyj} O, {Bodenheimer} P, et~al (1996) {Formation of the
  Giant Planets by Concurrent Accretion of Solids and Gas}. \icarus
  124(1):62--85. \doi{10.1006/icar.1996.0190}

\bibitem[{{Ratcliffe} et~al(2023){Ratcliffe}, {Minchev}, {Anders},
  {Khoperskov}, {Guiglion}, {Buck}, {Cunha}, {Queiroz}, {Nitschelm},
  {Meszaros}, {Steinmetz}, {de Jong}, {Nepal}, {Lane}, and
  {Sobeck}}]{Ratcliffe-23}
{Ratcliffe} B, {Minchev} I, {Anders} F, et~al (2023) {Unveiling the time
  evolution of chemical abundances across the Milky Way disc with APOGEE}.
  \mnras 525(2):2208--2228. \doi{10.1093/mnras/stad1573},
  {\href{https://arxiv.org/abs/2305.13378}{{arXiv:2305.13378}}} {[astro-ph.GA]}

\bibitem[{{Rauer} et~al(2024){Rauer}, {Aerts}, {Cabrera}, {Deleuil}, {Erikson},
  {Gizon}, {Goupil}, {Heras}, {Lorenzo-Alvarez}, {Marliani}, {Martin-Garcia},
  {Mas-Hesse}, {O'Rourke}, {Osborn}, {Pagano}, {Piotto}, {Pollacco},
  {Ragazzoni}, {Ramsay}, {Udry}, {Appourchaux}, {Benz}, {Brandeker},
  {G{\"u}del}, {Janot-Pacheco}, {Kabath}, {Kjeldsen}, {Min}, {Santos}, {Smith},
  {Suarez}, {Werner}, {Aboudan}, {Abreu}, {Acu{\~n}a}, {Adams}, {Adibekyan},
  {Affer}, {Agneray}, {Agnor}, {Aguirre B{\o}rsen-Koch}, {Ahmed}, {Aigrain},
  {Al-Bahlawan}, {Alcacera Gil}, {Alei}, {Alencar}, {Alexander},
  {Alfonso-Garz{\'o}n}, {Alibert}, {Allende Prieto}, {Almeida}, {Alonso
  Sobrino}, {Altavilla}, {Althaus}, {Alonso Alvarez Trujillo}, {Amarsi},
  {Ammler-von Eiff}, {Am{\^o}res}, {Andrade}, {Antoniadis-Karnavas},
  {Ant{\'o}nio}, {Aparicio del Moral}, {Appolloni}, {Arena}, {Armstrong},
  {Aroca Aliaga}, {Asplund}, {Audenaert}, {Auricchio}, {Avelino}, {Baeke},
  {Bailli{\'e}}, {Balado}, {Balestra}, {Ball}, {Ballans}, {Ballot}, {Barban},
  {Barbary}, {Barbieri}, {Barcel{\'o} Forteza}, {Barker}, {Barklem}, {Barnes},
  {Barrado Navascues}, {Barragan}, {Baruteau}, {Basu}, {Baudin}, {Baumeister},
  {Bayliss}, {Bazot}, {Beck}, {Bedding}, {Belkacem}, {Bellinger}, {Benatti},
  {Benomar}, {B{\'e}rard}, {Bergemann}, {Bergomi}, {Bernardo}, {Biazzo},
  {Bignamini}, {Bigot}, {Billot}, {Binet}, {Biondi}, {Biondi}, {Birch},
  {Bitsch}, {Bluhm Ceballos}, {B{\'o}di}, {Bogn{\'a}r}, {Boisse}, {Bolmont},
  {Bonanno}, {Bonavita}, {Bonfanti}, {Bonfils}, {Bonito}, {Bonomo},
  {B{\"o}rner}, {Boro Saikia}, {Borreguero Mart{\'\i}n}, {Borsa}, {Borsato},
  {Bossini}, {Bouchy}, {Bou{\'e}}, {Boufleur}, {Boumier}, {Bourrier}, {Bowman},
  {Bozzo}, {Bradley}, {Bray}, {Bressan}, {Breton}, {Brienza}, {Brito}, {Brogi},
  {Brown}, {Brown}, {Brun}, {Bruno}, {Bruns}, {Buchhave}, {Bugnet}, {Buldgen},
  {Burgess}, {Busatta}, {Busso}, {Buzasi}, {Caballero}, {Cabral}, {Calderone},
  {Cameron}, {Cameron}, {Campante}, {Canto Martins}, {Cara}, {Carone},
  {Carrasco}, {Casagrande}, {Casewell}, {Cassisi}, {Castellani}, {Castro},
  {Catala}, {Catal{\'a}n Fern{\'a}ndez}, {Catelan}, {Cegla}, {Cerruti},
  {Cessa}, {Chadid}, {Chaplin}, {Charpinet}, {Chiappini}, {Chiarucci},
  {Chiavassa}, {Chinellato}, {Chirulli}, {Christensen-Dalsgaard}, {Church},
  {Claret}, {Clarke}, {Claudi}, {Clermont}, {Coelho}, {Coelho}, {Cogato},
  {Colom{\'e}}, {Condamin}, {Conseil}, {Corbard}, {Correia}, {Corsaro},
  {Cosentino}, {Costes}, {Cottinelli}, {Covone}, {Creevey}, {Crida},
  {Csizmadia}, {Cunha}, {Curry}, {da Costa}, {da Silva}, {Dalal}, {Damasso},
  {Damiani}, {Damiani}, {Liduina das Chagas}, {Davies}, {Davies}, {Davies},
  {Davison}, {de Almeida}, {de Angeli}, {Cabral de Barros}, {de Castro
  Le{\~a}o}, {Brito de Freitas}, {de Freitas}, {De Martino}, {Renan de
  Medeiros}, {de Paula}, {de Plaa}, {De Ridder}, {Deal}, {Decin}, {Deeg},
  {Degl'Innocenti}, {Deheuvels}, {del Burgo}, {Del Sordo}, {Delgado-Mena},
  {Demangeon}, {Denk}, {Derekas}, {Desidera}, {Dexet}, {Di Criscienzo}, {Di
  Giorgio}, {Di Mauro}, {Diaz Rial}, {D{\'\i}az-Garc{\'\i}a}, {Dima},
  {Dinuzzi}, {Dionatos}, {Distefano}, {do Nascimento}, {Domingo}, {D'Orazi},
  {Dorn}, {Doyle}, {Duarte}, {Ducellier}, {Dumaye}, {Dumusque}, {Dupret},
  {Eggenberger}, {Ehrenreich}, {Eigm{\"u}ller}, {Eising}, {Emilio}, {Eriksson},
  {Ermocida}, {Isidoro Escate Giribaldi}, {Eschen}, {Estrela}, {Evans},
  {Fabbian}, {Fabrizio}, {Faria}, {Farina}, {Farinato}, {Feliz}, {Feltzing},
  {Fenouillet}, {Ferrari}, {Ferraz-Mello}, {Fialho}, {Fienga}, {Figueira},
  {Fiori}, {Flaccomio}, {Focardi}, {Foley}, {Fontignie}, {Ford}, {Fornazier},
  {Forveille}, {Fossati}, {de Marca Franca}, {da Silva}, {Frasca}, {Fridlund},
  {Furlan}, {Gabler}, {Gaido}, {Gallagher}, {Galli}, {Garcia}, {Garc{\'\i}a
  Hern{\'a}ndez}, {Garcia Munoz}, {Garc{\'\i}a-V{\'a}zquez}, {Garrido Haba},
  {Gaulme}, {Gauthier}, {Gehan}, {Gent}, {Georgieva}, {Ghigo}, {Giana}, {Gill},
  {Girardi}, {Giuliatti Winter}, {Giusi}, {Gomes da Silva}, {G{\'o}mez Zazo},
  {Gomez-Lopez}, {Isai Gonz{\'a}lez Hern{\'a}ndez}, {Gonzalez Murillo},
  {Gorius}, {Gouel}, {Goulty}, {Granata}, {Grenfell}, {Grie{\ss}bach},
  {Grolleau}, {Grouffal}, {Grziwa}, {Guarcello}, {Gueguen}, {Guenther},
  {Guilhem}, {Guillerot}, {Guiot}, {Guterman}, {Guti{\'e}rrez},
  {Guti{\'e}rrez-Canales}, {Hagelberg}, {Haldemann}, {Hall}, {Handberg},
  {Harrison}, {Harrison}, {Hasiba}, {Haswell}, {Hatalova}, {Hatzes}, {Haywood},
  {H{\'e}brard}, {Heckes}, {Heiter}, {Hekker}, {Heller}, {Helling},
  {Helminiak}, {Hemsley}, {Heng}, {Hermans}, {Hermes}, {Hidalgo Torres},
  {Hinkel}, {Hobbs}, {Hodgkin}, {Hofmann}, {Hojjatpanah}, {Houdek}, {Huber},
  {Huesler}, {Hui-Bon-Hoa}, {Huygen}, {Huynh}, {Iro}, {Irwin}, {Irwin},
  {Izidoro}, {Jacquinod}, {Emborg Jannsen}, {Janson}, {Jeszenszky}, {Jiang},
  {Jos{\'e} Jimenez Mancebo}, {Jofre}, {Johansen}, {Johnston}, {Jones},
  {Kallinger}, {K{\'a}lm{\'a}n}, {Kanitz}, {Karjalainen}, {Karjalainen},
  {Karoff}, {Kawaler}, {Kawata}, {Keereman}, {Keiderling}, {Kennedy},
  {Kenworthy}, {Kerschbaum}, {Kidger}, {Kiefer}, {Kintziger}, {Kislyakova},
  {Kiss}, {Klagyivik}, {Klahr}, {Klevas}, {Kochukhov}, {K{\"o}hler}, {Kolb},
  {Koncz}, {Korth}, {Kostogryz}, {Kov{\'a}cs}, {Kov{\'a}cs}, {Kozhura},
  {Krivova}, {Ku{\v{c}}inskas}, {Kuhlemann}, {Kupka}, {Laauwen}, {Labiano},
  {Lagarde}, {Laget}, {Laky}, {Lam}, {Lambrechts}, {Lammer}, {Lanza},
  {Lanzafame}, {Lares Martiz}, {Laskar}, {Latter}, {Lavanant}, {Lawrenson},
  {Lazzoni}, {Lebre}, {Lebreton}, {Lecavelier des Etangs}, {Leinhardt},
  {Leleu}, {Lendl}, {Leto}, {Levillain}, {Libert}, {Lichtenberg}, {Ligi},
  {Lignieres}, {Lillo-Box}, {Linsky}, {Scige Liu}, {Loidolt}, {Longval},
  {Lopes}, {Lorenzani}, {Ludwig}, {Lund}, {Sloth Lundkvist}, {Luri},
  {Maceroni}, {Madden}, {Madhusudhan}, {Maggio}, {Magliano}, {Magrin}, {Mahy},
  {Maibaum}, {Malac-Allain}, {Malapert}, {Malavolta}, {Maldonado}, {Mamonova},
  {Manchon}, {Mann}, {Mantovan}, {Marafatto}, {Marconi}, {Mardling}, {Marigo},
  {Marinoni}, {Marques}, {Marques}, {Marrese}, {Marshall}, {Mart{\'\i}nez
  Perales}, {Mary}, {Marzari}, {Masana}, {Mascher}, {Mathis}, {Mathur},
  {Mattiuci Figueiredo}, {Maxted}, {Mazeh}, {Mazevet}, {Mazzei}, {McCormac},
  {McMillan}, {Menou}, {Merle}, {Meru}, {Mesa}, {Messina}, {M{\'e}sz{\'a}ros},
  {Meunier}, {Meunier}, {Micela}, {Michaelis}, {Michel}, {Michielsen},
  {Michtchenko}, {Miglio}, {Miguel}, {Milligan}, {Mirouh}, {Mitchell},
  {Moedas}, {Molendini}, {Moln{\'a}r}, {Mombarg}, {Montalban}, {Montalto},
  {Monteiro}, {Morales}, {Morales-Calderon}, {Morbidelli}, {Mordasini},
  {Moreau}, {Morel}, {Morello}, {Morin}, {Mortier}, {Mosser}, {Mourard},
  {Mousis}, {Moutou}, {Mowlavi}, {Moya}, {Muehlmann}, {Muirhead}, {Munari},
  {Musella}, {Mustill}, {Nardetto}, {Nardiello}, {Narita}, {Nascimbeni},
  {Nash}, {Neiner}, {Nelson}, {Nettelmann}, {Nicolini}, {Nielsen}, {Niemi},
  {Noack}, {Noels-Grotsch}, {Noll}, {Norazman}, {Norton}, {Nsamba}, {Ofir},
  {Ogilvie}, {Olander}, {Olivetto}, {Olofsson}, {Ong}, {Ortolani}, {Oshagh},
  {Ottacher}, {Ottensamer}, {Ouazzani}, {Paardekooper}, {Pace}, {Pajas},
  {Palacios}, {Palandri}, {Palle}, {Paproth}, {Parro}, {Parviainen}, {Granado},
  {Passegger}, {Pastor-Morales}, {P{\"a}tzold}, {Gade Pedersen}, {Pena
  Hidalgo}, {Pepe}, {Pereira}, {Persson}, {Pertenais}, {Peter}, {Petit},
  {Petit}, {Pezzuto}, {Pichierri}, {Pietrinferni}, {Pinheiro}, {Pinsonneault},
  {Plachy}, {Plasson}, {Plez}, {Poppenhaeger}, {Poretti}, {Portaluri},
  {Portell}, {Frederico Porto de Mello}, {Poyatos}, {Pozuelos}, {Prada Moroni},
  {Pricopi}, {Prisinzano}, {Quade}, {Quirrenbach160}, {Rabanal Reina6},
  {Rabello Soares}, {Raimondo}, {Rainer}, {Ram{\'o}n Rod{\'o}n},
  {Ram{\'o}n-Ballesta}, {Ramos Zapata}, {R{\"a}tz}, {Rauterberg}, {Redman},
  {Redmer}, {Reese}, {Regibo}, {Reiners}, {Reinhold}, {Renie}, {Ribas},
  {Ribeiro}, {Pereira Ricciardi}, {Rice}, {Richard}, {Riello}, {Rieutord},
  {Ripepi}, {Rixon}, {Rockstein}, {Rodr{\'\i}guez}, {Rodr{\'\i}guez D{\'\i}az},
  {Rodriguez Garcia}, {Rodriguez-Gomez}, {Roehlly}, {Roig}, {Rojas-Ayala},
  {Rolf}, {Lysgaard R{\o}rsted}, {Rosado}, {Rosotti}, {Roth}, {Roth},
  {Rousseau}, {Roxburgh}, {Roy}, {Royer}, {Ruane}, {Rufini Mastropasqua}, {Ruiz
  de Galarreta}, {Russi}, {Saar}, {Saillenfest}, {Salaris}, {Salmon}, {Saltas},
  {Samadi}, {Samadi}, {Samra}, {Sanches da Silva}, {Andr{\'e}s S{\'a}nchez
  Carrasco}, {Santerne}, {Santoli}, {Santos}, {Sanz Mesa}, {Sarro},
  {Scandariato}, {Sch{\"a}fer}, {Schlafly}, {Schmider}, {Schneider}, {Schou},
  {Schunker}, {J{\"o}rg Schwarzkopf}, {Serenelli}, {Seynaeve}, {Shan},
  {Shapiro}, {Shipman}, {Sicilia}, {Sierra Sanmartin}, {Sigot}, {Silliman},
  {Silvotti}, {Simon}, {Simoyama Napoli}, {Skarka}, {Smalley}, {Smiljanic},
  {Smit}, {Smith}, {Smith}, {Snellen}, {S{\'o}dor}, {Sohl}, {Solanki},
  {Sortino}, {Sousa}, {Southworth}, {Souto}, {Sozzetti}, {Stamatellos},
  {Stassun}, {Steller}, {Stello}, {Stelzer}, {Stiebeler}, {Stokholm},
  {Storelvmo}, {Strassmeier}, {Str{\o}m}, {Strugarek}, {Sulis}, {{\v{S}}vanda},
  {Szabados}, {Szab{\'o}}, {Szab{\'o}}, {Szuszkiewicz}, {Talens}, {Teti},
  {Theisen}, {Th{\'e}venin}, {Thoul}, {Tiphene}, {Titz-Weider}, {Tkachenko},
  {Tomecki}, {Tonfat}, {Tosi}, {Trampedach}, {Traven}, {Triaud}, {Tr{\o}nnes},
  {Tsantaki}, {Tschentscher}, {Turin}, {Tvaruzka}, {Ulmer}, {Ulmer-Moll},
  {Ulusoy}, {Umbriaco}, {Valencia}, {Valentini}, {Valio}, {Valverde Guijarro},
  {Van Eylen}, {Van Grootel}, {van Kempen}, {Van Reeth}, {Van Zelst},
  {Vandenbussche}, {Vasiliou}, {Vasilyev}, {Vaz de Mascarenhas}, {Vazan}, {Vela
  Nunez}, {Nunes Velloso}, {Ventura}, {Ventura}, {Venturini}, {Trallero},
  {Veras}, {Verdugo}, {Verma}, {Vibert}, {Vicanek Martinez}, {Vida}, {Vigan},
  {Villacorta}, {Villaver}, {Villaverde Aparicio}, {Viotto}, {Vorobyov},
  {Vorontsov}, {Wagner}, {Walloschek}, {Walton}, {Walton}, {Wang}, {Waters},
  {Watson}, {Wedemeyer}, {Weeks}, {Weingril}, {Weiss}, {Wendler}, {West},
  {Westerdorff}, {Westphal}, {Wheatley}, {White}, {Whittaker}, {Wickhusen},
  {Wilson}, {Windsor}, {Winter}, {Lykke Winther}, {Winton}, {Witteck},
  {Witzke}, {Woitke}, {Wolter}, {Wuchterl}, {Wyatt}, {Yang}, {Yu}, {Zanmar
  Sanchez}, {Rosa Zapatero Osorio}, {Zechmeister}, {Zhou}, {Ziemke}, and
  {Zwintz}}]{Rauer-24}
{Rauer} H, {Aerts} C, {Cabrera} J, et~al (2024) {The PLATO Mission}. arXiv
  e-prints arXiv:2406.05447. \doi{10.48550/arXiv.2406.05447},
  {\href{https://arxiv.org/abs/2406.05447}{{arXiv:2406.05447}}} {[astro-ph.IM]}

\bibitem[{{Santos} et~al(2001){Santos}, {Israelian}, and {Mayor}}]{Santos-01}
{Santos} NC, {Israelian} G, {Mayor} M (2001) {The metal-rich nature of stars
  with planets}. \aap 373:1019--1031. \doi{10.1051/0004-6361:20010648},
  {\href{https://arxiv.org/abs/astro-ph/0105216}{{arXiv:astro-ph/0105216}}}
  {[astro-ph]}

\bibitem[{{Santos} et~al(2013){Santos}, {Sousa}, {Mortier}, {Neves},
  {Adibekyan}, {Tsantaki}, {Delgado Mena}, {Bonfils}, {Israelian}, {Mayor}, and
  {Udry}}]{Santos-13}
{Santos} NC, {Sousa} SG, {Mortier} A, et~al (2013) {SWEET-Cat: A catalogue of
  parameters for Stars With ExoplanETs. I. New atmospheric parameters and
  masses for 48 stars with planets}. \aap 556:A150.
  \doi{10.1051/0004-6361/201321286},
  {\href{https://arxiv.org/abs/1307.0354}{{arXiv:1307.0354}}} {[astro-ph.SR]}

\bibitem[{{Santos} et~al(2017{\natexlab{a}}){Santos}, {Adibekyan}, {Dorn},
  {Mordasini}, {Noack}, {Barros}, {Delgado-Mena}, {Demangeon}, {Faria},
  {Israelian}, and {Sousa}}]{Santos-17b}
{Santos} NC, {Adibekyan} V, {Dorn} C, et~al (2017{\natexlab{a}}) {Constraining
  planet structure and composition from stellar chemistry: trends in different
  stellar populations}. \aap 608:A94. \doi{10.1051/0004-6361/201731359},
  {\href{https://arxiv.org/abs/1711.00777}{{arXiv:1711.00777}}} {[astro-ph.EP]}

\bibitem[{{Santos} et~al(2017{\natexlab{b}}){Santos}, {Adibekyan}, {Figueira},
  {Andreasen}, {Barros}, {Delgado-Mena}, {Demangeon}, {Faria}, {Oshagh},
  {Sousa}, {Viana}, and {Ferreira}}]{Santos-17a}
{Santos} NC, {Adibekyan} V, {Figueira} P, et~al (2017{\natexlab{b}})
  {Observational evidence for two distinct giant planet populations}. \aap
  603:A30. \doi{10.1051/0004-6361/201730761},
  {\href{https://arxiv.org/abs/1705.06090}{{arXiv:1705.06090}}} {[astro-ph.EP]}

\bibitem[{{Schlaufman}(2018)}]{Schlaufman-18}
{Schlaufman} KC (2018) {Evidence of an Upper Bound on the Masses of Planets and
  Its Implications for Giant Planet Formation}. \apj 853(1):37.
  \doi{10.3847/1538-4357/aa961c},
  {\href{https://arxiv.org/abs/1801.06185}{{arXiv:1801.06185}}} {[astro-ph.EP]}

\bibitem[{Seabold and Perktold(2010)}]{Seabold-10}
Seabold S, Perktold J (2010) statsmodels: Econometric and statistical modeling
  with python. In: 9th Python in Science Conference

\bibitem[{{Sousa} et~al(2024){Sousa}, {Adibekyan}, {Delgado-Mena}, {Santos},
  {Rojas-Ayala}, {Barros}, {Demangeon}, {Hoyer}, {Israelian}, {Mortier},
  {Soares}, and {Tsantaki}}]{Sousa-24}
{Sousa} SG, {Adibekyan} V, {Delgado-Mena} E, et~al (2024) {SWEET-Cat: A view on
  the planetary mass-radius relation}. arXiv e-prints arXiv:2409.11965.
  \doi{10.48550/arXiv.2409.11965},
  {\href{https://arxiv.org/abs/2409.11965}{{arXiv:2409.11965}}} {[astro-ph.EP]}

\bibitem[{{Spiegel} et~al(2011){Spiegel}, {Burrows}, and {Milsom}}]{Spiegel-11}
{Spiegel} DS, {Burrows} A, {Milsom} JA (2011) {The Deuterium-burning Mass Limit
  for Brown Dwarfs and Giant Planets}. \apj 727(1):57.
  \doi{10.1088/0004-637X/727/1/57},
  {\href{https://arxiv.org/abs/1008.5150}{{arXiv:1008.5150}}} {[astro-ph.EP]}

\bibitem[{Virtanen et~al(2020)Virtanen, Gommers, Oliphant, Haberland, Reddy,
  Cournapeau, Burovski, Peterson, Weckesser, Bright, {van der Walt}, Brett,
  Wilson, Millman, Mayorov, Nelson, Jones, Kern, Larson, Carey, Polat, Feng,
  Moore, {VanderPlas}, Laxalde, Perktold, Cimrman, Henriksen, Quintero, Harris,
  Archibald, Ribeiro, Pedregosa, {van Mulbregt}, and {SciPy 1.0
  Contributors}}]{Virtanen-20}
Virtanen P, Gommers R, Oliphant TE, et~al (2020) {{SciPy} 1.0: Fundamental
  Algorithms for Scientific Computing in Python}. Nature Methods 17:261--272.
  \doi{10.1038/s41592-019-0686-2}

\bibitem[{Waskom(2021)}]{Waskom2021}
Waskom ML (2021) seaborn: statistical data visualization. Journal of Open
  Source Software 6(60):3021. \doi{10.21105/joss.03021},
  \urlprefix\url{https://doi.org/10.21105/joss.03021}

\end{thebibliography}

\newpage

\section*{Declarations}

\bmhead{Competing interests}
The authors have no relevant financial or non-financial interests to disclose.

\bmhead{Author contribution}
M.N. conducted most of the research/calculations and drafted the manuscript. V.A. provided guidance at key stages of the project and revisions of the manuscript. Both authors approved the final version for submission.

\bmhead{Data availability}
We used data from publicly available databases as outlined in the manuscript. The properties of the final sample of planets and their host stars is available at \href{https://zenodo.org/records/14331389}{Zenodo}.

\end{document}